\documentclass[aps,twocolumn]{revtex4}%
\usepackage{amsfonts}
\usepackage{amsmath}
\usepackage{amssymb}
\usepackage{graphicx}%
\begin{document}
\title{Influence of acceleration on multi-body entangled quantum states }
\author{Yongjie Pan}
\author{Baocheng Zhang}
\email{zhangbc.zhang@yahoo.com}
\affiliation{School of Mathematics and Physics, China University of Geosciences, Wuhan
430074, China}

\begin{abstract}
We study the influence of acceleration on the twin-Fock state which is a class
of specific multi-body entangled quantum state and was already realized
experimentally with high precision and sensitivity. We show that the
multi-body quantum entanglement can be increased with the acceleration,
consistent with the \textquotedblleft anti-Unruh effect\textquotedblright\ in
reference to the counterintuitive cooling previously pointed out for an
accelerated detector coupled to the vacuum. In particular, this kind of
entanglement increase can lead to the improvement of the phase sensitivity,
which provides a way to test the anti-Unruh effect in the future experiments.

\end{abstract}
\maketitle

\section{Introduction}

In 1976, Unruh discovered that an observer with uniform acceleration would
feel a thermal bath of particles in the Minkowski vacuum of a free quantum
field \cite{wgu76}, which implicates that the particle content of a quantum
field is observer dependent \cite{chm08}. This effect was put forward soon
after that Hawking discovered that black hole could emit thermal radiation
\cite{swh74} and could help to clarify some conceptual issues \cite{hhz19}
raised by black hole evaporation due to the equivalence. So the understanding
of the Unruh effect is also significant for Hawking radiation and the related
problems (i.e. information loss problems). In the past years, the Unruh effect
was digested and extended to many different situations (see the review
\cite{chm08} and references therein), but the observation of Unruh effect has
not been realized up to now, because of the pretty low Unruh temperature
$T={\hbar a}/{(2\pi ck_{B})}$ where $a$ is the proper acceleration of the
observer, ${\hbar}$ is the reduced Planck constant, $c$ is the speed of the
light, and ${k_{B}}$ is the Boltzman constant. The acceleration must be about
$10^{20}m/s^{2}$ in order to realize a photon bath at $1K$.

Although it was claimed that the Unruh effect is a direct result of quantum
field theory and does not require any experimental confirmation if the quantum
field theory is correct \cite{uw84}, there exists still some problems needed
to be clarified through experiments or observations, i.e. whether the
particles felt by the accelerated observers are real \cite{ccm16}, whether the
effect is applicable to the extended systems \cite{lbv19}, and even some
theoretical calculation implies that the possible inversion from Bose to Fermi
statistics for many-particle states observed by an accelerated observer
\cite{ltc17}. Thus, the experimental quest for the evidence of the Unruh
effect is necessary for the final confirmation. As well-known, the most
observational proposals are related to a model called the Unruh-DeWitt
detector \cite{bsd79}. Based on the model, it is found that a quantum system
consisting of such a detector uniformly accelerating in Minkowski vacuum sees
a thermal field and thus cause decoherence due to the coupling with the
thermal field. The first attempt is to observe such effect by the deexcitation
of the electron in storage rings by the thermal Unruh radiation \cite{bl83}.
Then some other possible detections related to proton decays \cite{vm01,sy03},
accelerated charges \cite{ssh06,oyz16}, neutrino oscillations \cite{blp20} and
the recent theoretical \cite{clv17} and observational \cite{lck19} methods
using Larmor radiation were proposed. In particular, an interesting
observation for Unruh radiation using quantum simulation in Bose-Einstein
condensates was reported, which is significant for the future research of the
dynamics of quantum many-body systems in a curved spacetime \cite{hfc19}.

On the other hand, the recent found anti-Unruh effect \cite{bmm16} states that
a particle detector in uniform acceleration coupled to the vacuum can cool
down with increasing acceleration under certain conditions, which is opposite
to the celebrated Unruh effect. Since the experiments are always made in the
range of finite length and time, it must distinguish the two situations of
Unruh and anti-Unruh effects carefully. An interesting way for this is to see
the change of quantum entanglement by acceleration. According to the previous
results \cite{fm05,amt06,ml09,mgl10,wj11,ses12,bfl12,ro15,dss15}, the quantum
entanglement would be degraded by the Unruh effect, which helps to establish
the general conclusion that entanglement is also observer dependent. In
particular, a recent calculation showed that the anti-Unruh effect can lead to
the increase for the quantum entanglement \cite{lzy18}, which might be
significant for the task of quantum information in large spatial or temporal
scale. In this paper, we will consider the influence of acceleration on the
spin squeezed states \cite{ku93,mws11} and the corresponding experimental
feasibility through the change of entanglement. Spin squeezed states have
attracted much attention due to their use in the measurement of the
correlation or entanglement among particles and in the improvement of
measurement precision in quantum metrology. We will focus on twin-Fock (TF)
states \cite{bh93} which can be seen as a kind of limit for spin squeezed
states and had been realized in a recent experiment with more than $10^{4}$
atoms \cite{lzy17}.

This paper is organized as follows. First, in section II we review the theory
about two-level Unruh-DeWitt (UDW) detector in Minkowski spacetime, and the
change of entanglement between two atoms for the Unruh and anti-Unruh effect.
This is followed in section III by the discussions on the influence of
acceleration on entanglement for TF states, where the spin squeezing parameter
is used to measure the change of entanglement. Then, when the atoms in the TF
state are accelerated, how the phase sensitivity is changed under the
background of the Ramsey interferometer is investigated in section IV.
Finally, we give a conclusion in section V. In this paper, we use units with
$c=\hbar=k_{B}=1$, except the part of analyzing the experimental feasibility
in section IV.

\section{The Unruh-DeWitt Model}

We start with the model of UDW detector in order to investigate the
interaction between accelerated atoms and vacuum. The detector, usually
considered as a point-like two-level quantum system or atom (as required in
this paper), consists of two quantum states, i.e. the ground $\left\vert
g\right\rangle $ and excited $\left\vert e\right\rangle $ states, which are
separated by an energy gap $\Omega$ while experiencing accelerated motion in a
vacuum field. But for the accelerated atom, the vacuum appears thermal due to
the Unruh effect, which will influence the state of the atom. This could be
described according to the following interaction Hamiltonian in a
($1+1$)-dimension model,
\begin{equation}
H_{I}=\lambda\chi\left(  \tau/\sigma\right)  \mu\left(  \tau\right)
\phi\left(  x\left(  \tau\right)  \right)  , \label{udwi}%
\end{equation}
where $\phi$ is a scalar field related to the vacuum in Minkowski spacetime
and interacts with the accelerated atom, $\lambda$ is the coupling strength,
$\tau$ is the atom's proper time along its trajectory $x\left(  \tau\right)
$, $\mu\left(  \tau\right)  $ is the atom's monopole momentum, and
$\chi\left(  \tau/\sigma\right)  $ is a switching function that is used to
control the interaction time scale $\sigma$. This can be easily generalized to
more complex situations such as a quantum oscillator \cite{bmm13}, as
confirmed with KMS conditions for thermal equilibrium \cite{gmr16}. For an
atom accelerating in a vacuum cavity, the evolution of the total quantum state
is determined perturbatively by the unitary operator up to first order given
by,
\begin{equation}
U=I-i\int d\tau H_{I}\left(  \tau\right)  +O\left(  \lambda^{2}\right)  .
\end{equation}

The atom is accelerated along the trajectory%
\begin{align}
t\left(  \tau\right)   &  =a^{-1}\sinh(a\tau),\nonumber\\
x\left(  \tau\right)   &  =a^{-1}\left(  \cosh(a\tau)-1\right)  , \label{itd}%
\end{align}
with the proper acceleration $a$. Note that the extra term $a^{-1}$ in the
expression of $x\left(  \tau\right)  $ is related to the initial condition and
only for convenience of the calculation below without changing the influence
of the acceleration. Thus, within the first-order approximation and in the
interaction picture, the evolution of the atom could be described by
\cite{bmm16},
\begin{align}
U\left\vert g\right\rangle \left\vert 0\right\rangle  &  =D_{0}\left(
\left\vert g\right\rangle \left\vert 0\right\rangle -i\eta_{_{0}}\left\vert
e\right\rangle \left\vert 1\right\rangle \right)  ,\nonumber\\
U\left\vert e\right\rangle \left\vert 0\right\rangle  &  =D_{1}\left(
\left\vert e\right\rangle \left\vert 0\right\rangle +i\eta_{_{1}}\left\vert
g\right\rangle \left\vert 1\right\rangle \right)  , \label{foe}%
\end{align}
where $k$ denotes the mode of the ($1+1$)-dimension scalar field with
(bosonic) annihilation (creation) operator $a_{k}$ ($a_{k}^{\dag}$),
$a_{k}\left\vert 0\right\rangle =0$ and $a_{k}^{\dag}\left\vert 0\right\rangle
=\left\vert 1_{k}\right\rangle $, and $D_{0,1}$ is the state normalization
factor. It is noted that the created state $\left\vert 1_{k}\right\rangle $ is
dependent on the wave vector $k$, so the coupling in Eq. (\ref{foe}) has to be
understood by writing $\eta_{_{0}}\left\vert 1\right\rangle =\lambda\int
dkI_{+,k}\left\vert 1_{k}\right\rangle $ and $\eta_{_{1}}\left\vert
1\right\rangle =\lambda\int dkI_{-,k}\left\vert 1_{k}\right\rangle $ where
$I_{\pm,k}$ is given as
\begin{equation}
I_{\pm,k}=\frac{1}{\sqrt{4\pi\omega}}\int_{-\infty}^{\infty}\chi\left(
\tau/\sigma\right)  \exp[\pm i\Omega\tau+i\omega t\left(  \tau\right)
-ikx\left(  \tau\right)  ]{d\tau}\text{.} \label{iii}%
\end{equation}
The notations $\eta_{_{0}}\left\vert 1\right\rangle $ and $\eta_{_{1}%
}\left\vert 1\right\rangle $ is inseparable, but in this paper we consider the
$\left\vert 1\right\rangle $ is the same for the two cases under the spirit of
single mode approximation \cite{am03,amm04}, and $\eta_{_{0}}$ and $\eta
_{_{1}}$ are related to the excitation and deexcitation probability of the
atom, i.e. $\left\vert \eta_{_{0}}\right\vert ^{2}=\sum_{k}\left\vert
\left\langle 1_{k},e\right\vert U^{(1)}\left\vert 0,g\right\rangle \right\vert
^{2}$ and $\left\vert \eta_{_{1}}\right\vert ^{2}=\sum_{k}\left\vert
\left\langle 1_{k},g\right\vert U^{(1)}\left\vert 0,e\right\rangle \right\vert
^{2}$ where $U^{(1)}=-i\int d\tau H_{I}\left(  \tau\right)  $.

It is worth pointing out that the change of the quantum state, i.e. the
transition probability, is dependent on the concrete parameters like the
interaction time scale $\sigma$ and the energy gap $\Omega$ \cite{bmm16}. In
particular, under some conditions, for example when the interaction timescale
is far away from the timescale associated to the reciprocal of the detector's
energy gap, the probability decreases as the acceleration or the Unruh
temperature increases, which makes the atom \textquotedblleft
feel\textquotedblright\ cooler instead of warm up expected by the Unruh
effect. This effect was called as anti-Unruh effect. Although the initial
discussion for the anti-Unruh effect is made in Ref. \cite{bmm16} for
accelerated detectors coupled to a massless scalar field either in a periodic
cavity or under a hard-IR momentum cutoff for the continuum, it has been shown
to represent a general stationary mechanism that can exist under a stationary
state satisfying Kubo-Martin-Schwinger (KMS) condition \cite{rk57,ms59,fjl16}
and is independent on any kind of boundary conditions \cite{bmm16,gmr16}.
Thus, like the Unruh effect, the anti-Unruh effect constitutes another new
phenomenon for the accelerated observers. Although the physically essential
reasons remain to be explored for their difference, some important elements,
like the interaction time, the detector's energy gap, the mass of the quantum
field, etc, had been pointed out to distinguish them operationally. Here we
consider massive field with e.g. $\omega=\sqrt{k^{2}+m^{2}}$ as in Ref.
\cite{gmr16} so that the anti-Unruh effect discussed will not be constrained
by the finite interaction time and its validity can be extended to situations
where the detector is switched on adiabatically over an infinite long time.
Without loss of generality, $m=1$ is used for all numerical calculations.

With that, the change of bipartite entanglement for two atoms was investigated
before \cite{lzy18}, in which the initial state is assumed to take the form
\begin{equation}
\left\vert \Psi_{i}\right\rangle =\left(  \alpha\left\vert g\right\rangle
_{A}\left\vert e\right\rangle _{B}+\beta\left\vert e\right\rangle
_{A}\left\vert g\right\rangle _{B}\right)  \left\vert 0\right\rangle
_{A}\left\vert 0\right\rangle _{B},
\end{equation}
with the complex coefficients satisfying $\left\vert \alpha\right\vert
^{2}+\left\vert \beta\right\vert ^{2}=1$. Here we consider the vacuum as in a
product state and thus the interaction between either one of two atoms and the
scalar field is independent of each other. This could help us to understand
the influence of acceleration on the quantum state of the atoms without the
disturbance of the complicated vacuum (i.e. it is regarded as an entangled
state) \cite{lzy18}. This means that the subscripts $A$ and $B$ in the vacuum
state $\left\vert \widetilde{0}\right\rangle \equiv$ $\left\vert
0\right\rangle _{A}\left\vert 0\right\rangle _{B}$ represents the locations
related to the atoms $A$ and $B$. For the case we consider, each atom is
independently \cite{itd} accelerating in the vacuum and has the same coupling
with the scalar field in its respective (spatial) place by the same process
presented in Eq. (\ref{foe}). When the two atoms are accelerated
simultaneously, the state becomes
\begin{align}
&  |\Psi_{f}\rangle=D_{0}D_{1}[\left(  \alpha\left\vert g\right\rangle
_{A}\left\vert e\right\rangle _{B}+\beta\left\vert e\right\rangle
_{A}\left\vert g\right\rangle _{B}\right)  \left\vert 0\right\rangle
_{A}\left\vert 0\right\rangle _{B}\nonumber\\
&  -i\left(  \alpha\eta_{_{1}}\left\vert g\right\rangle _{A}\left\vert
g\right\rangle _{B}+\beta\eta_{_{0}}\left\vert e\right\rangle _{A}\left\vert
e\right\rangle _{B}\right)  \left\vert 0\right\rangle _{A}\left\vert
1\right\rangle _{B}\nonumber\\
&  -i\left(  \beta\eta_{_{1}}\left\vert g\right\rangle _{A}\left\vert
g\right\rangle _{B}+\alpha\eta_{_{0}}\left\vert e\right\rangle _{A}\left\vert
e\right\rangle _{B}\right)  \left\vert 1\right\rangle _{A}\left\vert
0\right\rangle _{B}\nonumber\\
&  +\left(  \alpha\eta_{_{0}}\eta_{_{1}}\left\vert e\right\rangle
_{A}\left\vert g\right\rangle _{B}+\beta\eta_{_{0}}\eta_{_{1}}\left\vert
g\right\rangle _{A}\left\vert e\right\rangle _{B}\right)  \left\vert
1\right\rangle _{A}\left\vert 1\right\rangle _{B}],\hskip18pt \label{n2}%
\end{align}
where $\left\vert \widetilde{1}\right\rangle \equiv\left\vert 0\right\rangle
_{A}\left\vert 1\right\rangle _{B}\equiv$ $\left\vert 1\right\rangle
_{A}\left\vert 0\right\rangle _{B}$ represents the single-mode state from a
global perspective but $\left\vert 0\right\rangle _{A}\left\vert
1\right\rangle _{B}$ might be different from $\left\vert 1\right\rangle
_{A}\left\vert 0\right\rangle _{B}$ locally when the two atoms are separated
far apart, and $\left\vert \widetilde{2}\right\rangle \equiv\left\vert
1\right\rangle _{A}\left\vert 1\right\rangle _{B}$ represents the two-mode
state from a global perspective. It is necessary to keep the last term in Eq.
(\ref{n2}) in order to make the evolution in Eq. (\ref{foe}) intact formally,
since the so-called single-mode approximation in this paper is made for the
interaction between a single atom and the scalar field. When the two atoms
locates nearly at the same place, the forms $\left\vert 0\right\rangle
_{A}\left\vert 1\right\rangle _{B}$ and $\left\vert 1\right\rangle
_{A}\left\vert 0\right\rangle _{B}$ for the vacuum can be regarded as the same
and the two related terms in Eq. (\ref{n2}) can be combined into one, i.e.
$i(\alpha+\beta)(\eta_{_{1}}\left\vert g\right\rangle _{A}\left\vert
g\right\rangle _{B}+\eta_{_{0}}\left\vert e\right\rangle _{A}\left\vert
e\right\rangle _{B})\left\vert 0\right\rangle \left\vert 1\right\rangle $. We
consider the two atoms (or many atoms considered in the next section) staying
nearly in the same place in the whole process of acceleration, but the loss of
atoms due to acceleration is not considered in this paper.

The change of entanglement can be quantified by concurrence \cite{wkw98} which
is a widely used entanglement measure for bipartite mixed state. Concurrence
is defined by
\begin{equation}
C\left(  \rho\right)  =\max\{0,\lambda_{1}-\lambda_{2}-\lambda_{3}-\lambda
_{4}\},
\end{equation}
where $\lambda_{1}$, $\lambda_{2}$, $\lambda_{3}$, $\lambda_{4}$ are the
eigenvalues of the Hermitian matrix $\sqrt{\sqrt{\rho}\widetilde{\rho}%
\sqrt{\rho}}$ with $\widetilde{\rho}=\left(  \sigma_{y}\otimes\sigma
_{y}\right)  \rho^{\ast}\left(  \sigma_{y}\otimes\sigma_{y}\right)  $ the
spin-flipped state of $\rho$, $\sigma_{y}$ being the y-component Pauli matrix,
and the eigenvalues listed in decreasing order. For the case of two atoms
being accelerated, the change of entanglement can be calculated using
concurrence for the reduced density matrix $\rho_{AB}$ by tracing out the
scalar field from the final accelerated quantum state (\ref{n2}), which is
shown in the upper plot of Fig.1 with the initial state $\left\vert \Psi
_{i}\right\rangle $ are taken by $\alpha=\beta=\frac{1}{\sqrt{2}}$. The solid red
line represents the case of the anti-Unruh effect, as discussed in the
paragraph after Eq. (\ref{iii}). More detailed discussion and other cases for
different initial states refers to Ref. \cite{lzy18}. Since the experiment is
always made within certain timescale and using the certain energy gap, the
appearance of anti-Unruh effect is possible when the experiment is
implemented. Therefore, the experimental test must consider this point, for
which the increase of entanglement would also be the result of acceleration.

\begin{figure}
\centering
\includegraphics[width=1\columnwidth]{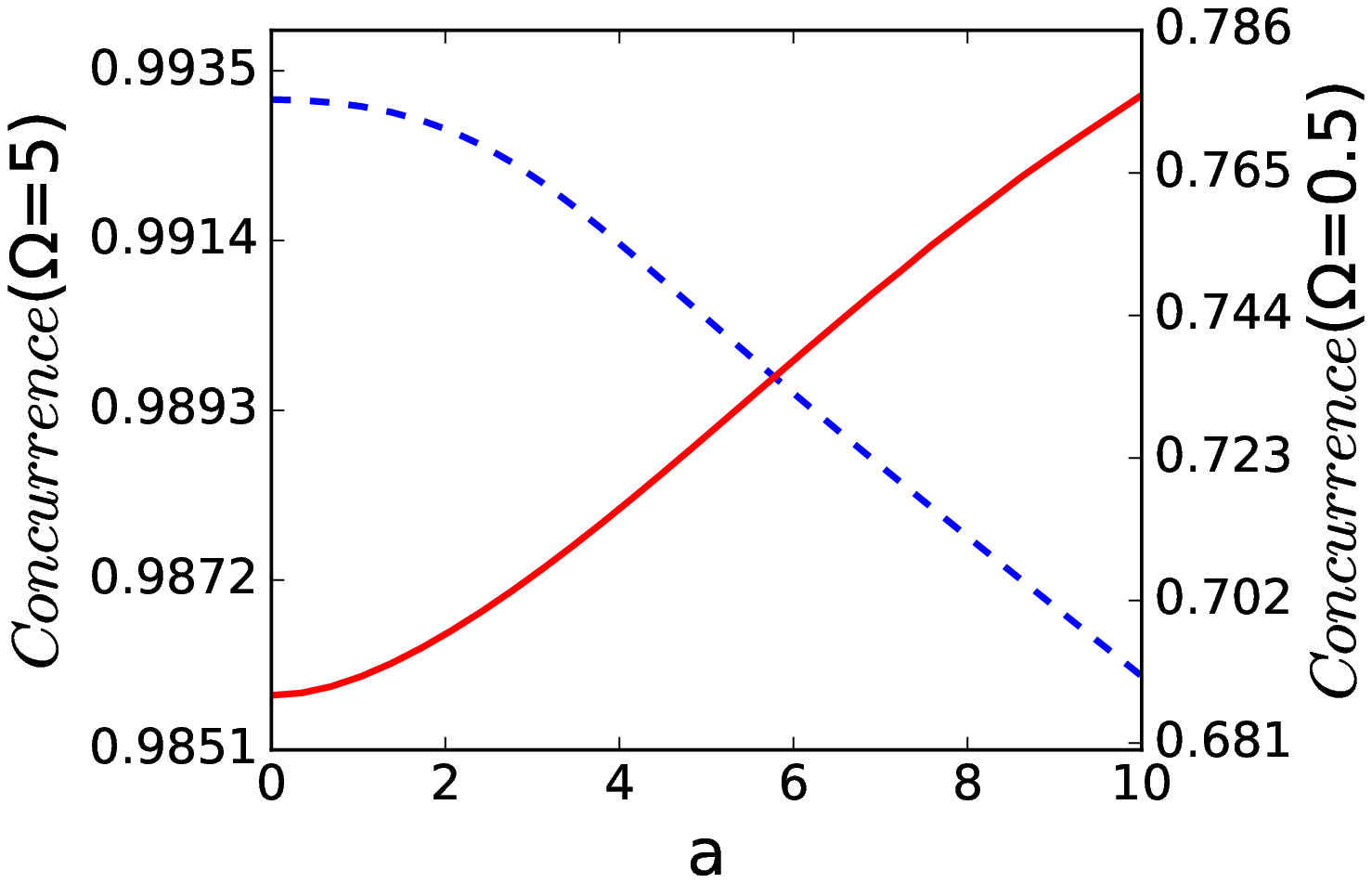}\\
\includegraphics[width=1\columnwidth]{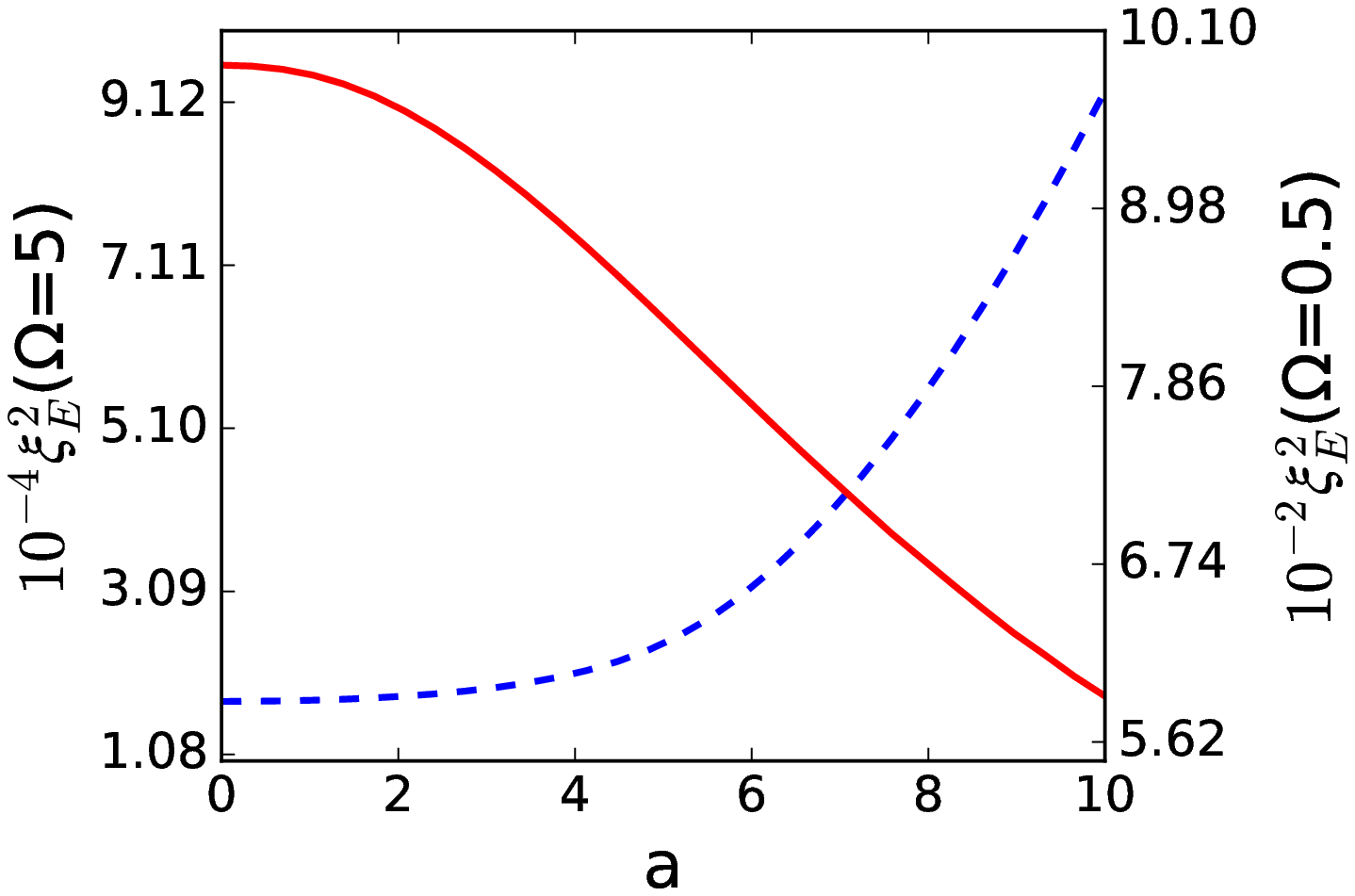}\\
\caption{(Color online) The concurrence (upper panel) and the spin squeezing
parameter (lower panel) as a
function of $a$ when two atoms are accelerated. The model parameters employed
are $\lambda=1$, $\sigma=0.4$. The solid red line denotes enhanced entanglement with
acceleration at $\Omega=0.5$ (referenced to the right vertical axis), while
the dashed blue line with respect to the left vertical axis is for the decreased
entanglement with acceleration at $\Omega=5$. We make the total atom number
$N=2$ for the right panel.}%
\label{Fig1}%
\end{figure}

\section{Twin-Fock State}

The previous section presents the UDW model and the change of entanglement
between two atoms in this model. Since it is not easy to implement the
corresponding experiment to observe the effect for two atoms, we now attempt
to apply it to the case of multi-body quantum states for multiple atoms being
simultaneously accelerated. Before that, it is noticed that the bipartite
quantum state for the maximal entangled atoms, $\left\vert \psi_{i}%
\right\rangle =\frac{1}{\sqrt{2}}\left(  \left\vert g\right\rangle
_{A}\left\vert e\right\rangle _{B}+\left\vert e\right\rangle _{A}\left\vert
g\right\rangle _{B}\right)  $, can be regarded as the simplest TF state. TF
state is one kind of Dicke states \cite{rhd54}. For a collection of $N$
identical (pseudo-) spin-1/2 particles, Dicke states can be expressed in Fock
space as $\left\vert \frac{N}{2}+m\right\rangle _{\uparrow}\left\vert \frac
{N}{2}-m\right\rangle _{\downarrow}$\ with ($\frac{N}{2}+m$) particles in
spin-up and ($\frac{N}{2}-m$) particles in spin-down modes for $m=-\frac{N}%
{2},-\frac{N}{2}+1,\cdots,\frac{N}{2}$. In particular, $m=0$ represents just
the TF state where the number of the particles is the same for each one of the
two spin states. On the other hand, Dicke states can be described by the
common eigenstate $|j,m\rangle$ of the collective spin operators $J^{2}$ and
$J_{z}$, with respective eigenvalues $j(j+1)$ and $m$. For the system
consisted of $N$ two-level atoms we will consider, the state $|j=\frac{N}%
{2},m\rangle$ indicates that ($j+m$) atoms are at the excited state
$|e\rangle$, ($j-m$) atoms are at the ground state $|g\rangle$, and it is the
TF state when $m=0$. $J_{z}=\frac{1}{2}\left(  n_{e}-n_{g}\right)  $
represents the difference of the number of atoms between excited ($n_{e}$) and
ground ($n_{g}$) states, and $J^{2}=\frac{N}{2}\left(  \frac{N}{2}+1\right)  $
is related to the total number of atoms. With this description, the state of
two atoms can be written as $\left\vert \psi_{i}\right\rangle =$ $|1,0\rangle
$, with which the average of the difference of the number of atoms,
$\left\langle J_{z}\right\rangle =0$.

When the two atoms are accelerated, according to the UDW model discussed in
last section, the state $\left\vert \psi_{i}\right\rangle $ will become
$\rho_{f}=Tr_{\phi}\left(  |\Psi_{f}\rangle\left\langle \Psi_{f}\right\vert
\right)  $ where $\alpha=\beta=\frac{1}{\sqrt{2}}$ are taken for the state
$|\Psi_{f}\rangle$, and $Tr_{\phi}$ indicates the calculation of tracing out
the part of the scalar field. Thus, the difference of the number of atoms
between excited and ground states is obtained as%
\begin{equation}
\left\langle J_{z}\right\rangle =Tr\left(  \rho_{f}J_{z}\right)  =D_{0}%
^{2}D_{1}^{2}\left(  \left\vert \eta_{0}\right\vert ^{2}-\left\vert \eta
_{1}\right\vert ^{2}\right)  ,
\end{equation}
where $Tr$ represents the trace of a matrix. The result means that the atom's
number at the excited state is not equal to that at the ground state,
different from the requirement of TF state, unless the probability of
transition from the ground state to the excited state equates the probability
for the inverse transition.

Now we extend this to the case of $N$ atoms with the initial TF state
$|j,0\rangle$. When all atoms are accelerated simultaneously, the TF state
becomes%
\begin{equation}
\rho_{t}=B_{0}^{2}|j,0\rangle\left\langle j,0\right\vert +\sum_{m=-N/2}%
^{N/2}\sum_{m^{\prime}=-N/2}^{N/2}B_{m}B_{m^{\prime}}^{\ast}|j,m\rangle
\left\langle j,m^{\prime}\right\vert , \label{dsm}%
\end{equation}
up to the normalization factor which is included in our numerical calculation.
$B_{0}^{2}=\sum_{k=0}^{N/2}\left[  \left(  C_{N/2}^{k}\right)  ^{4}\left(
D_{0}D_{1}\right)  ^{N}\left(  \eta_{_{0}}\eta_{_{1}}\right)  ^{2k}\right]  $,
$B_{m}=[\sum_{k=0}^{N/2-\left\vert m\right\vert }C_{N/2}^{k}C_{N/2}%
^{k+\left\vert m\right\vert }\left(  D_{0}D_{1}\right)  ^{N/2}\left(
\eta_{_{0}}\eta_{_{1}}\right)  ^{k}(\theta\left(  m\right)  \left(  -i\eta
_{0}\right)  ^{m}+\theta\left(  -m\right)  \left(  i\eta_{1}\right)
^{\left\vert m\right\vert })]$ in which the function $\theta\left(  x\right)
=1$ when $x>0$ and $\theta\left(  x\right)  =0$ otherwise, the star $\ast$ in
$B_{m^{\prime}}^{\ast}$ represents the complex conjugate, and $C_{n}^{r}%
=\frac{n!}{r!(n-r)!}$ denotes the combinatorial factor of choosing $r$ out of
$n$. When all atoms are accelerated, according to Eq. (\ref{foe}), the ground
state would change into the form $D_{0}\left(  \left\vert g\right\rangle
\left\vert 0\right\rangle -i\eta_{0}\left\vert e\right\rangle \left\vert
1\right\rangle \right)  $, and at the same time, the excited state $|e\rangle$
would change into the form $D_{1}\left(  \left\vert e\right\rangle \left\vert
0\right\rangle +i\eta_{1}\left\vert g\right\rangle \left\vert 1\right\rangle
\right)  $. Then, expanding these terms, recombining them and tracing the
vacuum state out give the expression (\ref{dsm}) of the final state after
acceleration. The parameter $B_{0}^{2}$ represents the probability of
remaining the original form of the TF state, which includes those cases that
if $l$ $\left(  0\leqslant l\leqslant\frac{N}{2}\right)  $ atoms are changed
from the ground states to the excited states, there must be other $l$ atoms
which are changed from the excited states to the ground states simultaneously.
Similarly, the parameter $B_{m}$ can be worked out by choosing the terms that
in every term either there are $m$ more excited states than ground states
(that is the case for $m>0$) or there are $m$ more ground states than excited
states (that is the case for $m<0$). No crossed terms like $|j,0\rangle
\left\langle j,m\right\vert $, because we consider the vacuum state including
the same number of photons as the same state no matter which atoms emitted
these photons. It is not difficult to confirm this for the cases $N=2$ and
$N=4$.

In order to quantify the change of entanglement in the process of accelerating
atoms that is initially in the TF state, we choose the spin squeezing
parameter \cite{mws11},
\begin{equation}
\xi_{E}^{2}=\frac{\underset{\overrightarrow{n}}{\min}\left[  \left(
N-1\right)  \left(  \Delta J_{\overrightarrow{n}}\right)  ^{2}+\left\langle
J_{\overrightarrow{n}}^{2}\right\rangle \right]  }{\left\langle J^{2}%
\right\rangle -N/2},
\end{equation}
which is rotationally invariant and related closely to entanglement. The
relation between spin squeezing and entanglement could be shown like that in
\cite{sdp01,tkb09}, in which the inequality%
\begin{equation}
\left(  N-1\right)  \left(  \Delta J_{\overrightarrow{n}}\right)
^{2}+\left\langle J_{\overrightarrow{n}}^{2}\right\rangle \geqslant
\left\langle J^{2}\right\rangle -N/2
\end{equation}
holds for any separable states, and the violation of this inequality indicates
entanglement. More related works refer to Ref. \cite{gt09}. If $\xi_{E}^{2}%
<1$, the state is spin squeezed and entangled. In particular, the smaller the
value of $\xi_{E}^{2}$, the more the entanglement will be, which can be seen
by comparing the upper and lower panels of Fig.1. Since the mean-spin direction
of Dicke states can be set along the z direction, we take z-direction as the
direction of $\overrightarrow{n}$ and the expression for the spin squeezing
parameter is written as
\begin{equation}
\xi_{E}^{2}=\frac{\left(  N-1\right)  \left(  \Delta J_{z}\right)
^{2}+\left\langle J_{z}^{2}\right\rangle }{\left\langle J^{2}\right\rangle
-N/2}. \label{sse}%
\end{equation}
For TF states, $\xi_{E}^{2}=0$ due to $\left\langle J_{z}^{2}\right\rangle
=\left\langle J_{z}\right\rangle =0$, which means that the initial TF state is
the most spin-squeezed and entangled state under the measure of $\xi_{E}^{2}$.

After acceleration, the TF state becomes $\rho_{t}$ described in Eq.
(\ref{dsm}). With this, we can calculate%
\begin{equation}
\left\langle J_{z}\right\rangle =Tr\left(  \rho_{t}J_{z}\right)
=\sum_{m=-N/2}^{N/2}m\left\vert B_{m}\right\vert ^{2}, \label{c1}%
\end{equation}
and
\begin{equation}
\left\langle J_{z}^{2}\right\rangle =Tr\left(  \rho_{t}J_{z}^{2}\right)
=\sum_{m=-N/2}^{N/2}m^{2}\left\vert B_{m}\right\vert ^{2}. \label{c2}%
\end{equation}
Thus, according to $\left(  \Delta J_{z}\right)  ^{2}=\left\langle J_{z}%
^{2}\right\rangle -\left\langle J_{z}\right\rangle ^{2}$, ones can calculate
$\xi_{E}^{2}$ by substituting these results (\ref{c1}) and (\ref{c2}) into Eq.
(\ref{sse}), which is presented in Fig.2 for different energy gaps (see also
the lower panel of Fig.1 for $N=2$). As seen, the anti-Unruh effect is
represented with the solid red line, and it shows that the entanglement increases
with the acceleration, as expected. It is noted that the entanglement at $a=0$
for the accelerated state (\ref{dsm}) is less than that for the initial
maximal entangled state due to the presence of switching function. This had
been pointed out before \cite{bmm16,bmm13} and its corresponding behavior in
entanglement was also presented clearly \cite{lzy18}. Moreover, we calculate
the change of $\xi_{E}^{2}$ with regard to the total number $N$ of atoms,
which is presented in Fig.3. It shows that entanglement with regard to the
initial entanglement or the change of entanglement increases when the number
$N$ increases for a given acceleration, no matter what energy gap is taken.
This makes the observation easier experimentally for the influence of
acceleration on the quantum state with a larger number. Although the trend of
the change is the same both for the Unruh and anti-Unruh effects, it appears
that the change of entanglement with $N$ atoms from the Unruh effect is more
violent than that from the anti-Unruh effect. This can be understood by
noticing that the atoms \textquotedblleft feel\textquotedblright\ hotter in
the case that the Unruh effect works than that the anti-Unruh effect works, as
seen for $a=10$ from Fig.2.

\begin{figure}[ptb]
\centering
\includegraphics[width=1\columnwidth]{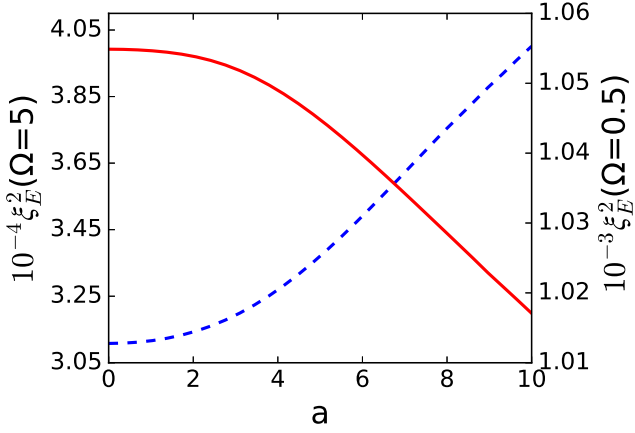} \caption{(Color online) The
spin squeezing parameter as a function of the acceleration $a$. We make the
total atom number $N=100$, and the other parameters are the same as in Fig.
1.}%
\label{Fig2}%
\end{figure}

\begin{figure}[ptb]
\centering
\includegraphics[width=1\columnwidth]{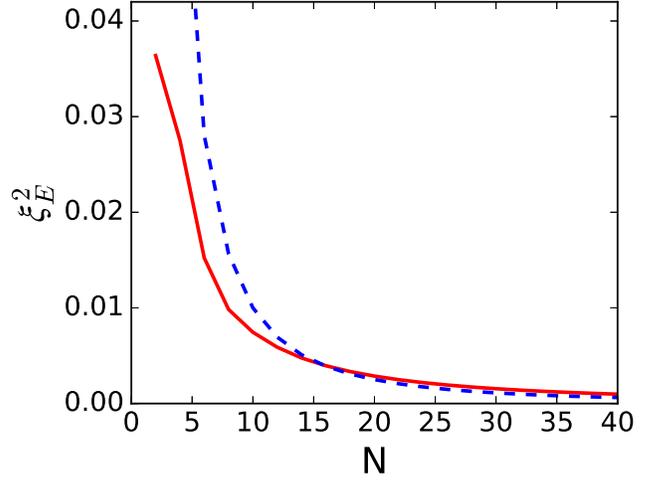} \caption{(Color online) The
spin squeezing parameter as a function of the total atom number $N$. The
acceleration is taken as $a=10$, and the other parameters are the same as in
Fig. 1. The diagram is drawn by linking the different points that are
calculated for every even number from $N=2$ to $N=40$.}%
\label{Fig3}%
\end{figure}

\section{Phase Sensitivity}

Since the change of spin squeezing or entanglement influences the phase
sensitivity of the measurement, in this section we will study the influence of
acceleration on the phase sensitivity and compare it with the present
experiment. In order to do this, we first give the general expressions, and
then compare the results for the TF state and its corresponding accelerated
state (\ref{dsm}).

Consider the Ramsey interferometer \cite{nr85,ymk86} with the initial input
state $\rho_{i}$, and the output state $\rho_{o}=U\rho_{i}U^{\dag}$ where
$U=\exp\left(  -i\theta J_{y}\right)  $ is the unitary operator for the
evolution and $\theta$ is the phase shift. According to the error propagation
formula \cite{mws11}, the phase sensitivity $\Delta\theta$ can be calculated
as%
\begin{equation}
\left(  \Delta\theta\right)  ^{2}=\frac{\left(  \Delta J_{z}^{2}\right)
_{o}^{2}}{\left\vert d\left\langle J_{z}^{2}\right\rangle _{o}/d\theta
\right\vert ^{2}},
\end{equation}
where the subscript $o$ denotes that the average is taken under the output
state. Using $UJ_{z}U^{\dag}=J_{z}\cos\theta-J_{x}\sin\theta$, it is easy to
calculate $\left\langle J_{z}^{2}\right\rangle _{o}=\left\langle J_{z}%
^{2}\right\rangle _{i}\cos^{2}\theta+\left\langle J_{x}^{2}\right\rangle
_{i}\sin^{2}\theta$ where the subscript $i$ denotes the average is taken under
the input state. It is seen that the phase shift can be deduced by measuring
$\left\langle J_{z}^{2}\right\rangle _{o}$ in the experiment. The
corresponding fluctuation of $J_{z}^{2}$ is $\left(  \Delta J_{z}^{2}\right)
_{o}^{2}=\left\langle J_{z}^{4}\right\rangle _{o}-\left\langle J_{z}%
^{2}\right\rangle _{o}^{2}=\left(  \Delta J_{z}^{2}\right)  _{i}^{2}\cos
^{4}\theta+\left(  \Delta J_{x}^{2}\right)  _{i}^{2}\sin^{4}\theta+V_{xz}%
\sin^{2}\theta\cos^{2}\theta$, where $V_{xz}=\left\langle \left(  J_{x}%
J_{z}+J_{z}J_{x}\right)  ^{2}\right\rangle _{i}+\left\langle J_{z}^{2}%
J_{x}^{2}+J_{x}^{2}J_{z}^{2}\right\rangle _{i}-2\left\langle J_{z}%
^{2}\right\rangle _{i}\left\langle J_{x}^{2}\right\rangle _{i}$. Thus, the
phase sensitivity becomes%
\begin{equation}
\left(  \Delta\theta\right)  ^{2}=\frac{\left(  \Delta J_{z}^{2}\right)
_{i}^{2}\cot^{2}\theta+\left(  \Delta J_{x}^{2}\right)  _{i}^{2}\tan^{2}%
\theta+V_{xz}}{4\left(  \left\langle J_{x}^{2}\right\rangle _{i}-\left\langle
J_{z}^{2}\right\rangle _{i}\right)  ^{2}}.
\end{equation}
When the phase shift satisfies $\tan^{2}\theta_{p}=\frac{\left(  \Delta
J_{z}^{2}\right)  _{i}}{\left(  \Delta J_{x}^{2}\right)  _{i}}$, the optimal
phase sensitivity is obtained as%
\begin{equation}
\left(  \Delta\theta\right)  _{P}^{2}=\frac{2\left(  \Delta J_{z}^{2}\right)
_{i}\left(  \Delta J_{x}^{2}\right)  _{i}+V_{xz}}{4\left(  \left\langle
J_{x}^{2}\right\rangle _{i}-\left\langle J_{z}^{2}\right\rangle _{i}\right)
^{2}}, \label{ops}%
\end{equation}
which is our main formula for investigating the change of phase sensitivity
due to the influence of acceleration on the spin squeezing or entanglement of
TF states.

For Dicke states $|j,m\rangle$, the optimal phase sensitivity occurs at
$\theta=0$ due to $\left(  \Delta J_{z}^{2}\right)  _{i}=0$. It is calculated
easily that $\left\langle J_{x}^{2}\right\rangle _{i}=\frac{1}{2}\left[
j\left(  j+1\right)  -m^{2}\right]  $, $\left\langle J_{z}^{2}\right\rangle
_{i}=m^{2}$, $V_{xz}=\frac{1}{2}\left(  4m^{2}+1\right)  \left[  j\left(
j+1\right)  -m^{2}\right]  -2m^{2}$. According to Eq. (\ref{ops}), the optimal
phase sensitivity for Dicke states is obtained as%
\begin{equation}
\left(  \Delta\theta\right)  _{PD}^{2}=\frac{\left(  4m^{2}+1\right)  \left[
j\left(  j+1\right)  -m^{2}\right]  -4m^{2}}{2\left[  j\left(  j+1\right)
-3m^{2}\right]  ^{2}}.
\end{equation}
When $m=j$, the result is $\frac{1}{2j}$ which is the standard quantum limit
and can be reached by the spin coherent state \cite{acg72}. When $m=0$, we
have%
\begin{equation}
\left(  \Delta\theta\right)  _{PD}^{2}=\frac{1}{2j\left(  j+1\right)  },
\end{equation}
which gives the phase sensitivity with $\sqrt{\frac{2}{N(N+2)}}$ approaching
the Heisenberg limit \cite{ps09}.

For the accelerated state in Eq. (\ref{dsm}), a direct but tedious calculation
within the approximation, $m,m^{\prime}<<j$ and $\left\vert B_{m}\right\vert
^{2}<<$ $B_{0}^{2}$ gives
\begin{align}
\left\langle J_{z}^{2}\right\rangle  &  =\sum_{m=-N/2}^{N/2}m^{2}\left\vert
B_{m}\right\vert ^{2},\nonumber\\
\Delta J_{z}^{2}  &  =\sqrt{\sum_{m=-N/2}^{N/2}m^{4}\left\vert B_{m}%
\right\vert ^{2}-\left(  \sum_{m=-N/2}^{N/2}m^{2}\left\vert B_{m}\right\vert
^{2}\right)  ^{2}},\nonumber\\
\left\langle J_{x}^{2}\right\rangle  &  \simeq\frac{1}{2}j\left(  j+1\right)
B_{0}^{2},\nonumber\\
\Delta J_{x}^{2}  &  \simeq\frac{B_{0}}{2\sqrt{2}}j\left(  j+1\right)
,\nonumber\\
V_{xz}  &  \simeq\frac{1}{2}j\left(  j+1\right)  \left[  1+\sum_{m=-N/2}%
^{N/2}\left\vert B_{m}\right\vert ^{2}\left(  4m^{2}+1\right)  \right]  .
\end{align}
Put these results into the Eq. (\ref{ops}), and the phase sensitivity is
obtained as%
\begin{align}
\left(  \Delta\theta\right)  _{PA}^{2}  &  \simeq\frac{1}{2j\left(
j+1\right)  }+\frac{\sqrt{2}B_{0}\Delta J_{z}^{2}}{2j\left(  j+1\right)
}\nonumber\\
&  +\frac{\sum_{m=-N/2}^{N/2}\left\vert B_{m}\right\vert ^{2}\left(
4m^{2}+1\right)  }{2j\left(  j+1\right)  }%
\end{align}
where the terms related to the summation are ignored in the denominator and we
have confirmed this approximation numerically. Fig.4 presents the behavior of
the phase sensitivity with regard to the acceleration. It is seen that when
the acceleration increases, the anti-Unruh effect can lead to the improvement
of the phase sensitivity. This is expected, since entanglement will increase
(this corresponds to the decrease of the squeezing parameter) when the quantum
state is accelerated in the case that the anti-Unruh effect works.

\begin{figure}[ptb]
\centering
\includegraphics[width=1\columnwidth]{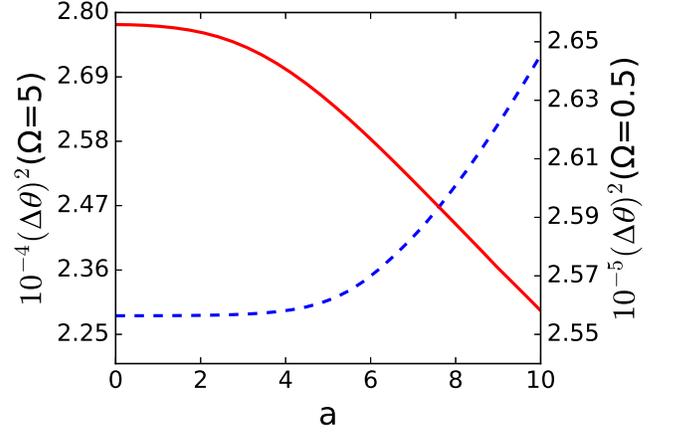} \caption{(Color online) The
phase sensitivity as a function of the acceleration $a$. We make the total
atom number $N=100$, and the other parameters are the same as in Fig. 1.}%
\label{Fig4}%
\end{figure}

\begin{figure}[ptb]
\centering
\includegraphics[width=1\columnwidth]{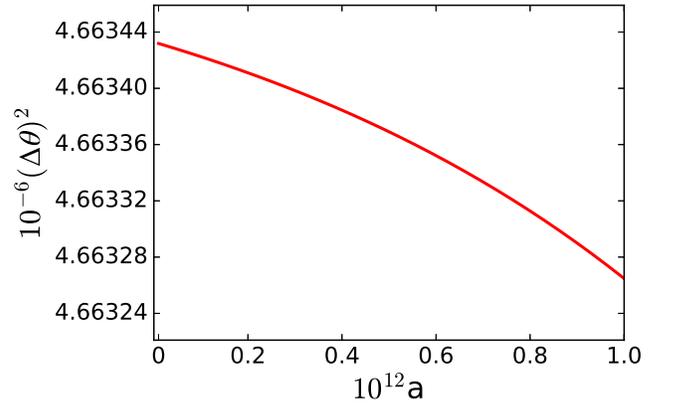} \caption{(Color online) The
phase sensitivity as a function of the acceleration $a$. We take the
parameters according to the experiment made in Ref. \cite{lzy17} with
$\lambda=1$, $\sigma=30$, $\Omega=2\pi$, and $N=10000$.}%
\label{Fig5}%
\end{figure}

Now we estimate the feasibility to test the effect through the corresponding
experiments. In the recent experiment that generates the TF state
\cite{lzy17}, the temperature is decreased to the level of $10^{-9}$ K that is
required to form Bose-Einstein condensates. Thus, in order to test the Unruh
or anti-Unruh effect, the acceleration has to reach the level of 10$^{10}$
m/s$^{2}$ at least, which is smaller than other experimental proposals
\cite{vm01,sy03,ssh06,oyz16,clv17} to test such effect, in which the
acceleration is more than 10$^{17}$ m/s$^{2}$. Together with the
experimentally allowable parameters, $\Omega\sim2\pi$ Hz, $N\sim10000$, it is
gotten that
\begin{equation}
\left(  \Delta\theta\right)  _{PA}^{2}\sim10^{-6},
\end{equation}
which is consistent with the present sensitivity in the experiment. This
doesn't mean that the Unruh or anti-Unruh effect can be tested in the
experiment instantly, because the required acceleration is still too large for
the practical implementation. However, such suggestion is promising by
reducing the acceleration through decreasing the experimental temperature or
increasing the number of atoms under the present sensitivity. It is also
feasible to reduce the acceleration by improving the sensitivity of
measurement by some means other than changing the temperature or the number of
atoms. In particular, for the anti-Unruh effect, it is noticed that the
influence of acceleration on the TF state could be extracted, even though the
temperature generated by the acceleration is lower than the temperature of
background, due to the fact that the thermal effect from the background cannot
lead to the increase of entanglement similar to the analysis made in Ref.
\cite{gmr16}. This is attractive for the future experiment with higher
sensitivity. Fig. 5 presents the possibility to realize the case of anti-Unruh
effect with the accelerating TF state.

\section{Conclusion}

In this paper, we revisit the influence of acceleration on quantum
entanglement and the possible test for this effect through accelerating one
class of experimentally feasible multi-body entangled quantum states. We have
calculated the change of the TF state due to acceleration and studied the
change of entanglement among atoms using the spin squeezing parameter as the
measurement of entanglement. It is shown that entanglement among atoms not
only decreases but also increases with the acceleration for the certain range
of the parameters. We have also compared the measurement of entanglement for
two atoms using concurrence and spin squeezing parameter respectively and
found that the same conclusion are obtained. In order to investigate the
feasibility of testing the effect from acceleration, we have calculated the
phase sensitivity of measurement using the distorted TF state due to
acceleration. It is interesting to note that the case for anti-Unruh effect
can appear for such accelerated states, which is favorable for the possible
future experiment since this effect is distinctive and different from that
coupled to a thermal environment directly by inertial observers \cite{gmr16}.

\section{Acknowledgement}

We would like to thank Li You, Qingyu Cai and Lingna Wu for their helpful
discussions. This work is supported by the NSFC grant Nos. 11654001 and 91636213.


\begin{thebibliography}{99}
\bibitem {itd}Noted that the two atoms move along the same (1+1) dimensional
trajectory in Eq. (\ref{itd}), but in fact the trajetories are different
because the initial positions for the two atoms could be different, i.e. their
initial coordinates y and z are different.
\end{thebibliography}
\end{document}